\documentclass[twoside,8pt]{article}
\usepackage{epsfig}

\newcommand{\pvec}{\vec{p}\,}

\newcommand{\pspace}{\int\limits_{|\pvec|\leq p_{F_{i}}}\!\!\!\!\!\! d^{3}p}
\newcommand{\fac}{\frac{\kappa}{(2\pi)^{3}}}

\newcommand{\be}{\begin{equation}}
\newcommand{\ee}{\end{equation}}
\newcommand{\bea}{\begin{eqnarray}}
\newcommand{\eea}{\end{eqnarray}}

\begin{document}


\title{ Bounds on the speed of sound in dense matter, and neutron star structure }

\author{Ch.C. Moustakidis, T. Gaitanos, Ch. Margaritis, and G.A. Lalazissis \\
Department of Theoretical Physics, Aristotle University of
Thessaloniki, \\ 54124 Thessaloniki, Greece }

\maketitle

\begin{abstract}

The accurate determination of the maximum mass of the neutron stars is one of the most important tasks in astrophysics. It is directly related to the identification of the black holes in the universe, the production of neutron stars from the supernovae explosion,  and the equation of state (EoS) of dense matter. However, not only the EoS is directly connected with neutron star masses, but also the speed of sound in dense matter is  a crucial quantity which  characterizes the stiffness of the EoS. The upper bound of the speed of sound  imposes strong constraints on the maximum mass of neutron stars. However, this upper bound remains still an open issue.  Recent observations, of binary neutron star systems,  offer the possibility of measuring  with high accuracy  both the mass and the tidal polarizability of the stars. We study possible effects of the upper bound of the speed of sound on  the upper bound of the mass and  the tidal polarizability. We conclude  that these kinds of measurements, combined  with recent observations of neutron stars with masses close to $2 M_{\odot}$, will provide  robust constraints on  the equation of state of hadronic matter at high densities.

\vspace{0.3cm}

PACS number(s):
26.60.-c, 21.30.Fe, 21.65.Cd, 26.60.Kp  \\

\end{abstract}

\section{Introduction}

The determination of the maximum mass of a neutron star (NS) (rotating and nonrotating) is one of the long-standing subjects  in astrophysics (for a comprehensive introduction dedicated to this problem see Ref.~\cite{Hartle-78}). In particular, the identification of a  black hole requires the knowledge of the maximum mass of a neutron star. The maximum neutron star mass is of considerable interest in the study of the production of neutron stars and black holes in the dynamics of supernovae explosion. Moreover, the experimental observations of neutron star masses have imposed strong constraints on the hadronic equation of state (EoS) of superdense matter (see also the references about the neutron star mass distribution~\cite{Coleman-015,Kiziltan-13}). The most famous examples are the recent discoveries of massive neutron stars with gravitational masses of $M=1.97\pm 0.04 \ M_{\odot}$ (PRS J1614-2230 \cite{Demorest-010}) and $M=2.01\pm 0.04 \ M_{\odot}$ (PSR  J0348+0432  \cite{Antoniadis-013}).

From a theoretical point of view, it is well known that the exact value of the maximum mass $M_{{\rm max}}$ of an NS depends strongly on the EoS of $\beta$-stable nuclear matter~\cite{Shapiro-83,Glendenning-2000,Haensel-07,Weinberg-72}. The EoS of compressed matter is somewhat well understood from studies of intermediate energy heavy-ion collisions. However, in such reactions the compressed hadronic fireball does not exceed  densities two or three times the saturation density. At higher densities the astrophysical observations on compact neutron stars are crucial for a better understanding of superdense hadronic matter, which is suppossed to exist in the core of these cosmic objects. Despite intensive investigations, the upper bound of neutron stars remains up to present uncertain~\cite{Rhoades-74,Kalogera-96,Koranda-97,Burgio-02,Schulze-06,Lattimer-010,Gandolfi-012,Bauswein-013,Chamel-015,Chamel-13,Dutra-016,Breu-016}. One possibility of proceeding  with an  estimate of $M_{{\rm max}}$ is based on the pioneering idea of Rhoades and  Ruffini~\cite{Rhoades-74}, where an optimum upper bound of mass of non-rotating neutron stars was derived using  a variational technique. In particular,  they considered  the most extreme EoS which produces the maximum mass solely  compatible with the following three conditions: (a) standard general-relativity equation of hydrostatic equilibrium, (b) Le Chattelier's principle, and (c) the principle of causality \cite{Rhoades-74}. The problem of finding an optimum bound on the mass of non-rotating neutron stars was considered also by other authors (see~\cite{Hartle-78} and references therein)  adopting similar assumptions to those of Ref.~\cite{Rhoades-74}.

So far, the theoretical assumptions for the construction of a neutron star EoS have been focused on the approximate  treatment of strongly interacting many-body systems. An issue was raised  recently concerning the high-density upper bound of the speed of sound. The speed of sound $v_s$, because of the causality, should not exceed that of light. However, as pointed out by Hartle \cite{Hartle-78},  causality arguments are not enough to allow a determination of the upper bound of the speed of sound.  Indeed, Weinberg~\cite{Weinberg-72} showed that the speed of sound is much less than the speed of light for  a cold nonrelativistic fluid.  It increases with temperature, but it does not exceed the value $c/\sqrt{3}$ at the limit of very high temperatures. The mentioned result seems to be unaffected when  electromagnetic forces are considered~\cite{Weinberg-72}.  However, this is still under debate. Lattimer~\cite{Lattimer-014} provided theoretical arguments that the causal limit is too extreme  because the highly compressed hadronic matter may convert asymptotically to free quark matter where the speed of sound  is $v_s=c/\sqrt{3} $.

Recently, Bedaque and Steiner \cite{Bedaque-015} have provided simple arguments that support the limit $c/\sqrt{3}$ in non-relativistic and/or weakly coupled theories.  This  was demonstrated in several classes of strongly coupled theories with gravity duals. The upper limit saturated only in conformal theories. The authors  pointed out that the existence of neutron stars with masses about two solar masses combined with the knowledge of the EoS of hadronic matter at low densities is not consistent with this  bound. Finally, it is worthy to mention that in a previous analysis
Olson~\cite{Olson-2000}  employed a phenomenological approach in the context of the kinetic theory,  to determine  an upper limit for the maximum mass of NS. In this approach the upper bound of the speed of sound appears to be less than the speed of light.

In view of the above investigations  we believe that  additional theoretical study is required. In this study,  the  effects of the upper bound of the speed of sound on the maximum mass of stars should be included in a comparison with the recent astronomical observations. The main motivation of the present paper is to study in detail the limiting cases of the upper bound of the speed of sound and their effects on the bulk  neutron star properties. We calculate maximum neutron star masses in relation to various scenarios for the upper bound of the speed of sound. We use a class of equation of states, which have been extensively employed in the literature and  mainly have the advantage to predict neutron star masses close or higher to the experimentally observed  value of $2 M_{\odot}$~\cite{Demorest-010,Antoniadis-013}.  In particular, we consider two upper bounds for the speed of sound, that is,  $v_s=c$ and $v_s=c/\sqrt{3}  $. Furthermore, a  phenomenological approach based on the kinetic theory is considered. This approach offers an additional independent theoretical prescription for the upper limit of the speed of sound in matter in relation to the maximum value of neutron star masses.

We also extend our study to  the analysis of the tidal polarizability (deformability), which  can be estimated experimentally.
The theoretical results for the tidal polarizability are discussed and analyzed in comparison with the corresponding observations  of the Advanced LIGO and the Einstein Telescope. We conclude, that the  accurate measurement of the tidal polarizabiliy $\lambda$ for a high mass neutron stars can be used to better understand the high-density stiffness of the EoS and the limit issue of the speed of sound in compressed matter.

The article is organized as follows: In Sec~II we briefly review the model for the maximum mass configuration. In Section~III we present the hadronic models for nuclear equation of state used in the present study.  In Section IV we analyze the tidal polarizability. The results are presented and discussed in Sec.~ V. The main conclusions of the present study, are presented in Sec~V.

\section{Nuclear equation of state and the maximum mass configuration}
It is known that no bounds can be determined for the mass of non-rotating neutron stars without some assumptions concerning the properties of neutron star matter \cite{Hartle-78}. In this study, following the work of Sabbadini and Hartle \cite{Sabbadini-1973,Hartle-77} we consider the  following four assumptions: (i) the matter of the neutron star is a perfect fluid described by a one-parameter equation of state between the pressure $P$ and the energy density ${\cal E}$, (ii) the energy density ${\cal E}$ is non-negative (because of the attractive character of gravitational forces), (iii) the matter is microscopically stable, which is ensured  by the conditions $P \geq 0$ and $dP/d{\cal E} \geq 0$ and (iv) below a critical baryon density $n_0$ the equation of state is well known.

From the above assumptions and the solution of the Tolman-Oppenheimer-Volkoff (TOV) equations it follows that the density and pressure decreases outward in the star. Furthermore, we introduce two regions for specifying more precisely the EoS. The radius $R_0$ at which the pressure is $P_0=P(n_0)$, divides the neutron star into two regions. The core, where $r \leq R_0$ and $n \geq n_0$ and the envelope where $r \geq R_0$ and $n \leq n_0$. The pressure $P$, the energy density ${\cal E}$ the density $\rho$ and the baryon density $n$ in nuclear mater are given as follows:
\begin{equation}
{\cal E}=n\left(\frac{}{} E+mc^2\right)=\rho c^2, \qquad  {\rm and} \quad P=n\frac{{\rm d}{\cal E}}{{\rm d}n}-{\cal E},
\label{es-1}
\end{equation}
where $E$ is the energy per baryon and $m$ is the nucleon mass.
The adiabatic speed of sound is defined as~\cite{Landau-87}
\begin{equation}
\frac{v_s}{c}=\sqrt{\left(\frac{\partial P}{\partial {\cal E}}\right)}_S,
\label{Vs}
\end{equation}
where $S$ is the entropy per baryon. In the present work we consider the following three  upper bounds  for the speed of sound:

\begin{enumerate}
\item  $\displaystyle \frac{v_s}{c} \leq 1$: causality limit from special relativity (see \cite{Hartle-78} and reference therein)

\item $\displaystyle \frac{v_s}{c} \leq \frac{1}{\sqrt{3}}$: from QCD and other theories (see \cite{Bedaque-015} and reference therein)

\item $\displaystyle \frac{v_s}{c}\leq \left(\frac{{\cal E}-P/3}{P+{\cal E}}\right)^{1/2}$: from relativistic kinetic theory (see \cite{Olson-2000} and reference therein)

\end{enumerate}

We construct the maximum mass configuration by considering the following structure for the neutron star EoS
\begin{eqnarray}
P({\cal E})&=&\left\{
\begin{array}{ll}
P_{crust}({\cal E}), \qquad  {\cal E} \leq {\cal E}_{\rm c-edge}       &          \\
\\
P_{NM}({\cal E}), \qquad  {\cal E}_{\rm c-edge} \leq {\cal E} \leq {\cal E}_{0}       &          \\
\\
\left(\frac{v_S}{c}  \right)^2\left({\cal E}-{\cal E}_{\rm c}  \right)+
P_{NM}({\cal E}_{0}), \qquad  {\cal E}_{0} \leq {\cal E}  .  &  \
                              \end{array}
                       \right.
\label{Basis-3}
\end{eqnarray}
According to Eq.~(\ref{Basis-3}), the EoS yielding the maximum mass of neutron stars,  is divided into three  regions. In particular, above the critical energy density ${\cal E}_0$ the EoS is maximally stiff with the speed of sound  $\sqrt{\left(\frac{\partial P}{\partial {\cal E}}\right)}_S$ fixed in the interval  $\left( 1/\sqrt{3}-1 \right)c$. In the  intermediate region ${\cal E}_{\rm c-edge} \leq {\cal E} \leq {\cal E}_{0}$ we employed a specific EoS which is used  for various nuclear models (see below for more details), while for ${\cal E} \leq {\cal E}_{\rm c-edge}$ we used the equation of Feynman, Metropolis and Teller~\cite{Feynman-49} and also of Baym, Bethe, and Sutherland~\cite{Baym-71}. The crust-core interface energy density ${\cal E}_{\rm c-edge} $, between the liquid core and the solid crust is determined  by employing the thermodynamical method~\cite{Moustakidis-010}. Actually, for energy densities lower than ${\cal E}_{\rm c-edge}$ the equation of state has negligible  effect on the maximum  mass configuration.  However, for the sake of completeness,  we use this additional crust structure in the calculations.

We  use the following notations and specifications for the results of the theoretical calculations: a) the case where the critical (fiducial) density is $n_0=1.5 n_s$ and for $n\geq n_0$ the speed of sound is fixed to the value $v_s=c$ (EoS/maxstiff), b) the case where  the fiducial density is $n_0=1.5 n_s$ and for $n\geq n_0$ the speed of sound is fixed to the value $v_s=c/\sqrt{3}$ (EoS/minstiff), and c) the case where the for $n\geq n_{c-crust}$ we simple employ the selected EoS without constraints (EoS/normal).

In any case, there are  very large numbers of possible combinations of EoS because the critical density $n_0$ and the upper bound of the speed of sound are not fixed from first principles in  the interval $(1.5-5)n_s$ and $(1/\sqrt{3}-1)c$, respectively. This results to a variety of EoSs lying between the two extreme cases: the EoS/maxstiff and the EoS/minstiff, which are described in more detail in the next section.

\section{The Nuclear Models}
In the present work we employed various relativistic and non-relativistic nuclear models, which are suitable to reproduce the bulk properties of nuclear matter at low densities, close to saturation density as well as the maximum observational neutron star mass (Refs.~\cite{Demorest-010,Antoniadis-013}). In particular, in each case,  the energy per particle of nuclear matter $E(I,n)$  is given as  a function of the baryonic number density $n$ and the asymmetry parameter $I=(n_n-n_p)/n=1-2x$ where $x$ is the proton fraction $n_p/n$. For $\beta$-stable matter, which is of interest here, the proton fraction $x$ is obtained from the condition,
\begin{equation}
\frac{\partial {\cal E}(u,x)}{\partial x}+\mu_e(u,x)=0,
\label{xpr-1}
\end{equation}
where ${\cal E}=E/n$ is the energy density. For an ultrarelativistic  degenerate electron gas  the chemical potential $\mu_e$ is given by
\begin{equation}
\mu_e(u,x)=\hbar c(3\pi^2x n_su)^{1/3},
\label{mu-el-1}
\end{equation}
where $n_s$ is the saturation density of the symmetric nuclear matter.
Below we provide a short description for each model separately.
\subsection{The MDI model }
The momentum-dependent interaction(MDI)  model used here, was already  presented and analyzed
in a previous paper \cite{Prakash-97}. The MDI model
is designed to reproduce the results of the microscopic
calculations of both nuclear and neutron-rich matter at zero
temperature and can be extended to finite temperature.
The energy per baryon  at $T=0$, is given by
\begin{eqnarray}
&&E(n,I)=\frac{3}{10}E_F^0u^{2/3}\left[(1+I)^{5/3}+(1-I)^{5/3}\right]+
\frac{1}{3}A\left[\frac{3}{2}-(\frac{1}{2}+x_0)I^2\right]u
\nonumber \\ &+&
\frac{\frac{2}{3}B\left[\frac{3}{2}-(\frac{1}{2}+x_3)I^2\right]u^{\sigma}}
{1+\frac{2}{3}B'\left[\frac{3}{2}-(\frac{1}{2}+x_3)I^2\right]u^{\sigma-1}}
 \label{e-T0}\\
&+&\frac{3}{2}\sum_{i=1,2}\left[C_i+\frac{C_i-8Z_i}{5}I\right]\left(\frac{\Lambda_i}{k_F^0}\right)^3
\left(\frac{\left((1+I)u\right)^{1/3}}{\frac{\Lambda_i}{k_F^0}}-
\tan^{-1} \frac{\left((1+
I)u\right)^{1/3}}{\frac{\Lambda_i}{k_F^0}}\right)\nonumber \\
&+&
\frac{3}{2}\sum_{i=1,2}\left[C_i-\frac{C_i-8Z_i}{5}I\right]\left(\frac{\Lambda_i}{k_F^0}\right)^3
\left(\frac{\left((1-I)u\right)^{1/3}}{\frac{\Lambda_i}{k_F^0}}-
\tan^{-1}
\frac{\left((1-I)u\right)^{1/3}}{\frac{\Lambda_i}{k_F^0}}\right),
\nonumber
\end{eqnarray}
where $u=n/n_s$, with $n_s$ denoting the
equilibrium symmetric nuclear matter density, $n_s=0.16$
fm$^{-3}$.   The parameters $A$, $B$, $\sigma$, $C_1$, $C_2$ and
$B'$ employed
to determine the properties of symmetric nuclear matter at saturation density $n_s$.
By suitably choosing the parameters $x_0$, $x_3$, $Z_1$, and
$Z_2$, it is possible to obtain different forms for the density
dependence of the symmetry energy as well as on the
value of the slope parameter $L$ and the value of the symmetry energy at the saturation density~\cite{Moustakidis-15}.  Actually, for each value of $L$ the density dependence  of the symmetry energy is adjusted so that the energy
of pure neutron matter is comparable with those of existing
state-of-the-art calculations~\cite{Moustakidis-15}.

\subsection{Momentum-dependent relativistic mean-field model}
The relativistic formulation of the nuclear matter problem is based on the well-known
quantum hadrodynamics (QHD)~\cite{Duerr:1956zz,Walecka:1974qa,Serot:1997xg,Vretenar-05}.
The QHD describes the nuclear matter in terms of
Dirac spinors for the nucleons, which interact through the exchange of mesons in the spirit
of the one-boson-exchange model~\cite{Serot:1984ey}. According to the covariant structure these
virtual
mesons are characterized by a Lorentz-scalar, iso-scalar $\sigma$, a Lorentz-vector,
iso-scalar $\omega$ and a Lorentz-vector, iso-vector $\rho$ fields~\cite{Vretenar-05}. The $\sigma$ and
$\omega$ exchange fields are responsible for the binding mechanism of ordinary nuclear
matter, while the $\rho$ field is required for the description of isospin asymmetric matter.

Here we adopt an RMF approach with non-linear derivative interactions, the so-called
non-linear derivative (NLD) model; see Ref.~\cite{Gaitanos:2012hg,Gaitanos:2015xpa}
for details. It is suitable for
applications in systems beyond saturation density, because it contains explicitly the
momentum dependence of the interaction. The NLD approach reproduces the bulk nuclear matter
properties and, simultaneously, the empirical momentum dependence of the in-medium
nucleon~\cite{Gaitanos:2012hg} and antinucleon~\cite{Gaitanos:2015xpa} potentials.

The energy density in NLD is obtained from the time-like $00$-component of the
energy-momentum tensor and it reads
\begin{equation}
{\cal E} =
\sum_{i=p,n} \fac \pspace \, E(\pvec) +
\frac{1}{2}
\left(
m_{\sigma}^{2}\sigma^{2} + 2U(\sigma) - m_{\omega}^{2}\omega^{2}- m_{\rho}^{2}\rho^{2}
\right)
\label{ED-NLD}
\,.
\end{equation}
The first term in Eq.~(\ref{ED-NLD}) is the kinetic contribution, while the other
terms appear because of the in-medium interaction, mediated by the three virtual
$\sigma, \omega$ and $\rho$ mesons with masses $m_{\sigma}$, $m_{\omega}$ and
$m_{\rho}$, respectively. The additional contribution $U(\sigma)$ includes the
conventional non-linear self-interactions of the $\sigma$ field, as proposed originally
by Boguta~\cite{Boguta:1977xi}. Equation~(\ref{ED-NLD}) for the energy density looks similar as the
familiar expressions of the usual RMF models. However, the non-linear effects are induced
by the NLD regulators. As discussed in Ref.~\cite{Gaitanos:2012hg}, the momentum
dependence regulates
the high momentum behavior of the mesonic source terms and affects the dispersion relation,
from which the energy $E=E(\vec{p}\,)$ is calculated. The NLD equation of state is soft
at densities just above the saturation density, but it becomes stiff at higher densities.
It reproduces the recent observations of two-solar mass neutron stars.

\subsection{The HLPS model}
Recently, Hebeler {\it et al.}~\cite{Hebeler-010,Hebeler-013} performed microscopic calculations based on chiral effective field theory interactions  to constrain the properties of neutron-rich matter below nuclear densities. It explains the massive neutron stars of $M=2 M_{\odot}$.  In this model the energy per particle is given by~\cite{Hebeler-013} (hereafter HLPS model)
\begin{eqnarray}
\frac{E(u,x)}{T_0}&=&\frac{3}{5}\left(x^{5/3}+(1-x)^{5/3}  \right)(2u)^{2/3}-\left[(2\alpha-4\alpha_L)x(1-x)+\alpha_L\right]u\nonumber\\
&+&
\left[(2\eta-4\eta_L)x(1-x)+\eta_L\right]u^{\gamma},
\label{ED-HLPS}
\end{eqnarray}
where $T_0=(3\pi^2n_0/2)^{2/3}\hbar^2/(2m)=36.84$ MeV.
The corresponding pressure is given by
\begin{eqnarray}
\frac{P(u,x)}{n_0T_0}&=&\frac{2}{5}\left(x^{5/3}+(1-x)^{5/3}  \right)(2u)^{5/3}-\left[(2\alpha-4\alpha_L)x(1-x)+\alpha_L\right]u^2\nonumber\\
&+&
\gamma\left[(2\eta-4\eta_L)x(1-x)+\eta_L\right]u^{\gamma+1}.
\label{Pres-HLPS}
\end{eqnarray}
The parameters $\alpha$, $\eta$, $\alpha_L$ and $\eta_L$ are determined by combining   the saturation properties of symmetric nuclear matter and the microscopic calculations for neutron matter~\cite{Hebeler-010,Hebeler-013}. The parameter $\gamma$ is used to adjust the values of the incompessibility $K$ and influences the range on the values of the symmetry energy and its density derivative.
%
\subsection{The H-HJ model}
Heiselberg and Hjorth-Jensen~\cite{Heiselberg-2000,Heiselberg-1999} adopted the following  simple form for the energy  per particle in nuclear matter
(hereafter H-HJ  model)
\begin{equation}
E=E_0u \frac{u-2-d}{1+d u}+S_ou^{\gamma}(1-2x)^2,
\label{H-H-eos-1}
\end{equation}
where $E_0=-15.8 \ {\rm MeV}$. The parameters $d$ and $\gamma$ are fixed to the EoS of Akmal et al.~\cite{Akmal-1998}, both for pure neutron matter and symmetric nuclear matters. However, the high-density extrapolation is questionable, because the EoS of Akmal {\it et al}.~\cite{Akmal-1998} becomes superluminal.

\subsection{The Skyrme models}

Finally, we also perform calculations using the well-known Skyrme parametrization. The energy per baryon of asymmetric matter is given by~\cite{Chabanat-97,Farine-97}
\begin{eqnarray}
E(n,I)&=&\frac{3}{10}\frac{\hbar
^2c^2}{m}\left(\frac{3\pi^2}{2}\right)^{2/3}n^{2/3}F_{5/3}(I)+\frac{1}{8}t_0n\left[2(x_0+2)-(2x_0+1)F_2(I)\right]
\nonumber \\
&+&\frac{1}{48}t_3n^{\sigma+1}\left[2(x_3+2)-(2x_3+1)F_2(I)\right] \\
&+& \frac{3}{40}\left(\frac{3\pi^2}{2}\right)^{2/3}n^{5/3}
\left[\frac{}{}\left(t_1(x_1+2)+t_2(x_2+2)\right)F_{5/3}(I)
\right.
\nonumber\\
&+&\left.\frac{1}{2}\left(t_2(2x_2+1)-t_1(2x_1+1)\right)F_{8/3}(I)\right],
\nonumber
\end{eqnarray}
where $\displaystyle F_m(I)=\frac{1}{2}\left[(1+I)^m+(1-I)^m
\right] $. The parametrization is  given in  Refs.~\cite{Chabanat-97,Farine-97}.

\subsection{The relativistic kinetic theory and the upper bound of the speed of sound}
The relativistic kinetic theory also predicts an upper bound of  the speed of sound, which differs from unity.  Douchin and Haensel found that the condition $v_s \leq c$ is not sufficient to respect causality in neutron star matter~\cite{Douchin-2001}. Additional causality constraints can be derived from relativistic kinetic theory. This was achieved by Olson using a Grad method of moments~\cite{Olson-2000,Israel-76,Israel-79}. He modeled viscous and thermal dissipation in neutron star matter and determined an upper limit for the maximum mass of neutron stars~\cite{Olson-2000}. According to the relativistic kinetic theory, the EoS must satisfy various conditions.  Equilibrium fluid states should be thermodynamically stable and perturbations relative to equilibrium should propagate causally via hyperbolic equations~\cite{Olson-2000}.

To summarize, in  the Israel-Stewart  theory~\cite{Israel-76,Israel-79} the thermodynamical  variables characterizing  the fluid must also satisfy a set of conditions  for the fluid to be thermodynamically stable, causal, and hyperbolic. The functional forms for the relation times and the viscous-thermal coupling strengths, which are the basic ingredients of the conditions mentioned above,  have been calculated  in the noninteracting degenerate Fermi gases~\cite{Israel-79,Olson-89}.  In the low temperature limit $1/kT\longrightarrow \infty$  the conditions are given by the following inequalities (for more details of the derivations of the inequalities see Ref.~\cite{Olson-89})
\begin{equation}
{\cal E}\geq 0, \qquad P\geq 0, \qquad (P+{\cal E}) \left(\frac{v_s}{c}\right)^2\geq 0, \qquad \left(\frac{v_s}{c}\right)^2\leq \frac{{\cal E}-P/3}{P+{\cal E}},
\qquad P\leq 3 {\cal E}.
\label{kin-cond-all}
\end{equation}
The conditions in Eqs.~(\ref{kin-cond-all}) impose stringent constraints on the high-density equation of state and thus, stringent constraints on the maximum neutron star mass. The requirement of these conditions implies that the maximally stiff equation of state fulfills the following expression:
\begin{equation}
\left(\frac{v_s}{c}\right)^2= \frac{{\cal E}-P/3}{P+{\cal E}}.
\label{equal-kin-4}
\end{equation}
Now, considering that
\[ \left(\frac{v_s}{c}\right)^2=\frac{d P}{d {\cal E}}\]
and also that, at temperature $T=0$, the pressure is given by the expression
\[P=n^2\frac{d ({\cal E}/n)}{dn}=n\frac{d{\cal E}}{d n} -{\cal E},  \]
Eq.~~(\ref{equal-kin-4}), after some simple algebra,   leads to the differential equation,
\begin{equation}
n^2\frac{d^2{\cal E}}{dn^2}+\frac{n}{3}\frac{d{\cal E}}{dn}-\frac{4}{3}{\cal E}=0.
\label{Difer-1}
\end{equation}
Equation~(\ref{Difer-1}) can be easily solved and leads to
\begin{equation}
{\cal E}(n)={\cal C}_1n^{a_1}+{\cal C}_2n^{a_2}, \qquad  a_1=(1+\sqrt{13})/3, \ a_2=(1-\sqrt{13})/3,
\label{En-kin-1}
\end{equation}
and also
\begin{equation}
P(n)={\cal C}_1n^{a_1}(a_1-1)+{\cal C}_2n^{a_2}(a_2-1).
\label{Pres-kin-1}
\end{equation}
Solving the system of Eqs.~(\ref{En-kin-1}) and ~(\ref{Pres-kin-1}) we found that
\begin{equation}
{\cal C}_1=\left(\frac{(2+\sqrt{13}){\cal E}(n)+3P(n)}{2\sqrt{13}}\right)n^{-\frac{1}{3}(1+\sqrt{13})},
\label{C1-1}
\end{equation}
and
\begin{equation}
{\cal C}_2=-\left(\frac{(2-\sqrt{13}){\cal E}(n)+3P(n)}{2\sqrt{13}}\right)n^{-\frac{1}{3}(1-\sqrt{13})}.
\label{C2-1}
\end{equation}
The values of the constants ${\cal C}_1$ and ${\cal C}_2$ are determined with the help of the critical density $n_0$. Now, the total equation of state, suitably to describe the maximum mass configuration of a neutron matter, is given by the ansatz [see again Eq.~(\ref{Basis-3})]
\begin{eqnarray}
P(n)&=&\left\{
\begin{array}{ll}
P_{crust}(n), \qquad  n \leq n_{\rm c-edge}       &          \\
\\
P_{NM}(n), \qquad  n_{\rm c-edge} \leq n \leq n_{\rm 0}       &          \\
\\
{\cal C}_1n^{a_1}(a_1-1)+{\cal C}_2n^{a_2}(a_2-1)
, \qquad   n_{\rm 0} \leq n .  &  \
                              \end{array}
                       \right.
\label{Kinetic-eos}
\end{eqnarray}
For the equation of sate  in the interval $n_{\rm c-edge} \leq n \leq n_{\rm 0}$ we employed the MDI model with $L=110$ MeV. However, the results, especially for low values of the fiducial density $n_0$ are model independent. It is worth pointing out  that the condition $\left(\frac{v_s}{c}\right)^2\leq \frac{{\cal E}-P/3}{P+{\cal E}}$ is well satisfied everywhere.

\section{Tidal Polarizability}


Gravitational waves from the final stages of inspiraling binary neutron stars are
expected to be one of the most important sources
for ground-based gravitational wave detectors \cite{Flanagan-08,Hinderer-08,Damour-09,Hinderer-10,Postnikov-010,Fattoyev-13,Lackey-015}. The masses of the component of the system will be
determined with moderate accuracy, especially if the neutron stars are slowly spinning,
during the early stage of the evolution. Flanagan and Hinderer \cite{Flanagan-08} have recently pointed out
that tidal effects are also potentially measurable during the early part of the evolution
when the waveform is relatively clean. The tidal fields induce quadrupole moments on the neutron stars.
The response of the neutron star is described by the dimensionless so-called Love number $k_2$,
which depends on the neutron star structure and consequently on the mass and the EoS of the nuclear matter.
The tidal Love numbers $k_2$ is obtained from the ratio of the induced quadrupole moment $Q_{ij}$ to the applied tidal field $E_{ij}$:
\begin{equation}
Q_{ij}=-k_2\frac{2R^5}{3G}E_{ij}\equiv \lambda E_{ij},
\label{Love-1}
\end{equation}
where $R$ is the neutron star radius  and $\lambda=2R^5k_2/3G$ is the tidal polarizability. The tidal Love number $k_2$ is given by \cite{Flanagan-08,Hinderer-08}
\begin{eqnarray}
k_2&=&\frac{8\beta^5}{5}\left(1-2\beta\right)^2\left[2-y_R+(y_R-1)2\beta \right] \nonumber \\
&\times&
\left[\frac{}{} 2\beta \left(6  -3y_R+3\beta (5y_R-8)\right) \right. \nonumber \\
&+& 4\beta^3 \left.  \left(13-11y_R+\beta(3y_R-2)+2\beta^2(1+y_R)\right)\frac{}{} \right.\nonumber \\
&+& \left. 3\left(1-2\beta \right)^2\left[2-y_R+2\beta(y_R-1)\right] {\rm ln}\left(1-2\beta\right)\right]^{-1},
\label{k2-def}
\end{eqnarray}
where $\beta=GM/Rc^2$ the compactness parameter. The tidal Love number $k_2$ depends on the compactness parameter $\beta$  and the quantity $y_R$. Actually, $y_R$ is determined by solving the following differential equation for $y$
\begin{equation}
r\frac{dy(r)}{dr}+y^2(r)+y(r)F(r)+r^2Q(r)=0, \qquad y(0)=2,\quad y_R\equiv y(R)
\label{D-y-1}
\end{equation}
where $F(r)$ and $Q(r)$ are functionals of ${\cal E}(r)$, $P(r)$ and $M(r)$  defined as~\cite{Hinderer-10,Postnikov-010}
\begin{equation}
F(r)=\left[ 1- \frac{4\pi r^2 G}{c^4}\left({\cal E} (r)-P(r) \right)\right]\left(1-\frac{2M(r)G}{rc^2}  \right)^{-1},
\label{Fr-1}
\end{equation}
and
\begin{eqnarray}
r^2Q(r)&=&\frac{4\pi r^2 G}{c^4} \left[5{\cal E} (r)+9P(r)+\frac{{\cal E} (r)+P(r)}{\partial P(r)/\partial{\cal E} (r)}\right]
\left(1-\frac{2M(r)G}{rc^2}  \right)^{-1}  \\
&-&6\left(1-\frac{2M(r)G}{rc^2}  \right)^{-1} \nonumber \\
&-&\frac{4M^2(r)G^2}{r^2c^4}\left(1+\frac{4\pi r^3 P(r)}{M(r)c^2}   \right)^2\left(1-\frac{2M(r)G}{rc^2}  \right)^{-2}.
\nonumber
\label{Qr-1}
\end{eqnarray}
Equation (\ref{D-y-1})  must be integrated self-consistently  with the TOV equations using the boundary conditions $y(0)=2$, $P(0)=P_c$ and $M(0)=0$. The solution of the TOV equations provides the mass $M$ and radius $R$ of the neutron star, while the corresponding solution of the differential Eq.~(\ref{D-y-1}) provides the value of $y_R=y(R)$. This together with the quantity $\beta$ are the  basic ingredients  of the tidal Love number $k_2$ [Eq.~(\ref{k2-def})].

In addition, the combined tidal effects of two neutron stars in a circular orbits  are given by a weighted average of the quadrupole
responses~\cite{ Flanagan-08,Hinderer-10},
\begin{equation}
\tilde{\lambda}=\frac{1}{26}\left[\frac{m_1+12m_2}{m_1}\lambda_1+\frac{m_2+12m_1}{m_2}\lambda_2  \right],
\label{aver-tital}
\end{equation}
where $\lambda_1=\lambda_1(m_1)$ and $\lambda_2=\lambda_2(m_2)$ are the tidal deformabilities of the two neutron stars and  $M=m_1+m_2$ the total mass. The symmetric mass ratio is defined as $h=m_1m_2/M^2$. The dependence  of $\lambda$ on the mass $m$ can be calculated  for individual neutron stars. However, as pointed out in Ref.~\cite{Hinderer-10}, the universality of the neutron star EoS allows  one to predict the tidal phase contribution for a given binary system  from each EoS. In this case the weighted average $\tilde{\lambda}$ is usually  plotted as a function of chirp mass ${\cal M}=(m_1m_2)^{3/5}/M^{1/5}$ for various values of the ratio $h$.

\section{Results and Discussion}
We  perform a systematic study of the speed of sound effects on the hadronic equation of state adopting the nuclear models, for $\beta$-stable matter, described in Chapter 3. Strangeness degrees of freedom (hyperons) were not considered in the present study. It should be noted  that the inclusion of hyperons may lead to a softening of the hadronic EoS. However,  to keep the  present work as clear as possible, we left out the strangeness effects for a separate work.
We concentrate  our study mainly on an estimate of the maximum mass of neutron stars, by taking into account the various constraints on the upper bound of the speed of sound. We also compare the results with a few recent observations. In particular, we analyze the tidal polarizability and its possible dependence  on the hadronic EoS. Indeed, as we shall see, this observable is sensitive to the EoS and, hence, can be used as a sensitive probe for the maximum mass configuration.

We start the discussion with Fig.~1. It  shows the radius-mass relation of neutron stars using various EoS without any restriction on the speed of sound (except the relativistic one). One can see, that all hadronic models can reproduce the recent observation of two-solar massive neutron stars. In general, the stiffer EoS (at high densities) the higher  the maximum neutron star mass. This effect is clearly presented Fig.~1 for the HLPS EoS  of the soft, intermediate, and stiff case. This is a general feature. It arises from the fact, that for a soft EoS the matter can be easier compressed relative to a stiff one. Thus, less gravitational energy is needed for a stable neutron star configuration.

Before starting to analyze the effects of the speed of sound limits on the EoS, we show in Fig. 2 the density dependence of this quantity for the various EoSs used here. It is obvious that almost all the  EoSs are causal even for high values of the pressure (the only exception is the case HLPS (stiff) where the $v_c$ exceed the $c$ for relative low pressure). However, it is worth   mentioning that in all hadronic models, used in the present study, the speed of sound  $v_s$ reaches the bound limit $c/\sqrt{3} $ at relative low values of the pressure (for $P \leq 100\ {\rm MeV}\cdot {\rm fm}^{-3}$). This feature has dramatic effect on the maximum mass configuration. This is now discussed in Fig. 3, where the radius-mass relation for the different nuclear models is shown including the various scenarios for the upper limit of the speed of sound. The neutron star configurations with the five selected EoSs in the normal case (no constraints on $v_s$ except the conventional $ v_s<c$) are also shown for comparison. It is seen that the upper bound limit of the speed of sound imposes essential changes to the neutron star structure. The higher the limit after the fiducial density, the stiffer the corresponding EoS. This results to a higher value for the maximum neutron star mass. By setting the upper limit to $v_s=c/\sqrt{3}$ the stiffness of the EoS weakens at higher densities and consequently the neutron star mass  reduces to lower values. Certainly, the results will depend on the onset of the critical density. However, the effects of the upper bound of $v_s$ on the neutron star structure are too strong. Therefore, the upper limit issue of the in-medium speed of sound should be seriously considered in such studies.

To further clarify the critical density dependence on $M_{{\rm max}}$, we display  in Fig.~4 the dependence of the maximum mass for the chosen  EoS, on  the fiducial density $n_0$. We considered three upper bounds for the speed of sound: $v_s=c$,  $v_s=c/\sqrt{3}$, and the bound originated from the kinetic theory [see Eq.~(\ref{equal-kin-4})]. First, one sees an overall reduction of the neutron star mass with increasing critical density. Using the density behavior of the $v_s=c/\sqrt{3}$ constraint in the calculations, the neutron star mass first decreases and then approaches a constant  value, which is characteristic for each EoS. It is remarkable that in all cases the neutron star mass drops below the experimental value of two solar masses (the only exception is the stiff case of the HLPS model).  Therefore, the assumption of $v_s=c/\sqrt{3}$ value as the upper limit for the speed of sound in compressed matter would exclude particular EOSs which contradict  with  recent astrophysical observation of massive neutron stars. Our results are similar to those of Bedaque and Steiner~\cite{Bedaque-015}. In that  paper the authors found that the existence of neutron stars with masses around two solar masses, combined with a specific EoS of hadronic matter at low densities, is inconsistent   with the upper limit  $v_s=c/\sqrt{3} $~\cite{Bedaque-015}.

On the other hand, when the causality limit $v_s=c$ is imposed, the upper bound on the maximum mass significantly increases as is well known from previous studies and the relevant predictions (see Refs.~\cite{Lattimer-010,Chamel-015} and references therein). Our calculations are in agreement with the observations of the massive neutron stars even for high values of the fiducial densities $n_0$. The predictions of the kinetic theory for the  $M_{\rm max}$ are lower but  close to those of the causality limit and also in accordance with the observations.
It is noted that recently an upper bound of neutron star masses was obtained from an analysis of  short gamma-ray bursts~\cite{Lawrence-15}.
According to these studies, most short gamma-ray bursts are
produced by the coalescence of two neutron stars, and if the merger remnant collapses quickly, then the upper mass
limit is constrained tightly. Assuming that the rotation of the merger remnant is limited only by mass-shedding, then the maximum gravitational mass of a nonrotating neutron star is $M_{\rm max}=(2-2.2) \ M_{\odot}$~\cite{Lawrence-15}.

In a similar analysis the physical properties of the compact remnant  in neutron star mergers have been studied in Ref.~\cite{Fryer-15} by combining  the results of Newtonian merger calculations and equation-of-state studies. Furthermore, the authors of Ref.~\cite{Fryer-15}, using population studies,  determined the distribution of these compact remnants to compare with the observations. It turned out that black
hole cores are formed quickly only for equations of state that predict maximum non-rotating neutron star masses below
$M_{\rm max}=(2.3 - 2.4) \ M_{\odot}$. All these recent analyses, although dealing  with the problem of the upper bound of neutron star mass in different ways,  predict  as an absolute  upper limit for the  maximum neutron star masses  $M_{{\rm max}}\geq 2 M_{\odot}$.  This prediction is  in accordance with the recent observations~\cite{Demorest-010,Antoniadis-013}. Obviously, the theoretical predictions of the present work, with the imposed upper bound of  $v_s=c/\sqrt{3}$, fail to reproduce the absolute upper limits of $(2.3 - 2.4) \ M_{\odot}$. In any case, only additional observations will  clarify further the problem of the $M_{\rm max}$ and consequently to provide  additional constraints for the upper mass limit.

In Fig.~5, we display the dependence of the $M_{{\rm max}}$ on  the upper bound of the speed of sound and for three different values of the fiducial density. The dependence is of the form $M_{{\rm max}}=a(v_s/c)^b M_{\odot}$. The values of the parameters $a$ and $b$, in each case, have
been determined by the least-squares fit method.  We consider as an example the case of the MDI model. The calculations show that the situation is similar to  the other EoSs.   As expected, for  the lower values of the fiducial density $n_0$ the values of  $M_{{\rm max}}$ increase more rapidly with $v_s$. Obviously the  $M_{{\rm max}}-v_s$ dependence becomes less pronounced  and for low  $v_s$ values close to the results of the EoS/normal case.

We propose now an additional approach to investigate the upper bound of $v_s$. In particular, we calculate for the EoSs used in this work and for the various maximum mass configurations, the corresponding values of the tidal polarizability.  It is well known that the influence of the star's internal structure on the  waveform is characterized by the value of the tidal polarizability $\lambda$. This quantity $\lambda$ is actually a measure of the star's quadrupole deformation in response to the companion's perturbing tidal field~\cite{Hinderer-10}. It was  found that $\lambda$ is sensitive to the details of the equation of state. The Love number $k_2$ shows a strong dependence on the compactness parameter $\beta$ (for high values of $\beta$ decreases very fast) as well as on the value of $y_R$ which depends also on the internal structure of the neutron stars.   Furthermore, the tidal polarizability exhibits very strong dependence on the radius $R$ and consequently on the details of the equation of state at low and high values of the baryons density. In particular, the maximum mass and consequently the compactness parameter are sensitive to the high density region of the EoS. As confirmed in studies by Lattimer and Prakash~\cite{Lattimer-01}, the radius of neutron stars with masses close to $1.4 M_{\odot}$ depends on the pressure near the saturation density. This leads us  to believe that the measure of the tidal polarizability may imposes  constraints  related to the maximum mass configuration and also the maximum mass of a neutron star involved in a binary neutron star system.

Figure~6 shows the dependence  of the Love number $k_2$ on the mass of a neutron star for the five selected EoSs and in comparison with the corresponding maximum mass configurations results (for the upper bounds $v_s=c, c/\sqrt{3}$). The quantity $k_2$ is obviously model dependent especially for medium values of the mass ($0.5-1.5 M_{\odot}$). Close to the value of $1 M_{\odot}$ there  is a deviation by a factor $\sim 2-3$ between the various models. Actually, $k_2$ is strongly dependent on the compactness parameter $\beta$ [see Eq.~(\ref{k2-def})]. For example, in the HLPS model, the low values of $k_2$ in the intermediate mass region of $0.5-1 M_{\odot}$ are because of the large values of the compactness $\beta$ (see Fig. 3). The interplay between the upper bound of the speed of sound and the stiffness of the EoS thus influences in a non-trivial way the Love number $k_2$. In particular, the values of $k_2$ are relatively insensitive to the form of the EoS, when the $c/\sqrt{3}$ limit is imposed at the fiducial density $n_0=1.5n_s$ and  for neutron star masses above one solar mass ($M > 1 M_{\odot}$). However, when using the conventional causality limit ($v_s < c$), the stiffness of the EoS becomes more pronounced with significant effects on the tidal Love number $k_2$. In fact, the quantity $k_2$ increases by a factor of two or even more relative to the results with the $c/\sqrt{3}$ limit. Moreover, the corresponding maximum mass configuration in the causality limit ($v_s<c$ at $n_0=1.5n_s$) gives finite $k_2$-values even for high neutron star masses greater that two solar masses ($M>2 M_{\odot}$). This feature is very different with the one obtained with  the calculations using the EoS/normal case for each model.

The tidal polarizability is an important quantity, as it can be deduced from observations on neutron star binary systems. This is shown in Fig. 7.  The notation is the same as in Fig.~6. The {\it signature } of the maximum mass configuration on the values of $\lambda$ is obvious. In particular, we found that $\lambda$ takes a wide range of values ($\lambda\sim (1-5)\times 10^{36}$ gr cm$^2$ s$^2$) for the employed EoS (EoS/normal case). Because $\lambda$ is sensitive to the neutron star radius, this quantity is directly affected by the EoS. An EoS leading to large neutron star radii will also give high values for the tidal polarizability $\lambda$ (and vice versa). The constraints of the upper bound on the speed of sound (EoS/minstiff) lead to a non-negligible increase of $\lambda$ for high values of neutron star mass. However, in the EoS/maxstiff case the corresponding increase  of $\lambda$ is substantial, compared to the EoS/normal case. Moreover, in this case the values of $\lambda$ remain measurable even for very high values of the mass.   This behavior results from the strong dependence of $\lambda$ on the radius $R$. Specifically, according to Fig.~3 the increase  of the upper bound on the speed of sound influences  significantly the maximum mass configuration in two ways. First, a dramatic increase of the upper bound of $M_{\rm max}$. Second,  the neutron star radius is significantly increased. A radius increase by $10 \%$ leads already to a rise of the tidal polarizability $\lambda$ by $60 \%$.

In the same figure, the ability to measure  the tidal polarizability from the Advanced LIGO and the Einstein Telescope is indicated. The region of possible  observations with the Advanced LIGO is indicated by the unshaded region. The Einstein Telescope has larger ability and will be able to measure the tidal polarizability in the unshaded and light shaded region (see also Ref.~\cite{Hinderer-10}). It it obvious that the Advanced LIGO is able to measure $\lambda$ and to impose constraints on the EoS only for low values of neutron stars. However, as was  pointed out in Ref.~\cite{Hinderer-10} the Einstein Telescope  increases significantly the region of sensitivity  and will offers the opportunity to give more stringent constraints on the equation of state both for low and high values of the density.

Note that the Einstein Telescope will be able to measure $\lambda$ even for neutron stars with a masses up to $2.5 \ M_{\odot}$ and consequently to constrain the stifness of the equation of state. To be more precise, from these observations one will be able to test the upper bound $v_s=c/\sqrt{3}$. The simultaneous measurements of neutron star masses $M$ and tidal polarizabilities $\lambda$ will definitely help to better clarify  the stiffness limits of the equation of state.
These features are shown in our calculations. For instance, the model  within the EoS/normal case predicts values of the tidal polarizability $\lambda$, which are out of the detection region of the Einstein Telescope (see also the studies in Refs.~\cite{Hinderer-10,Postnikov-010,Fattoyev-13}). On the other hand, the calculations with the EoS/minstiff model lead to $\lambda$ values, which are just near that sensitivity region. The EoS/maxstiff results show a clear observable signature. In particular, for intermediate mass neutron stars $(1-2 M_{\odot})$ with large values for the tidal polarizability $\lambda$ the upper bound $v_s=c/\sqrt{3}$ seems to be violated. In view of the above analysis we conjecture that it is possible with the third-generations  detectors to examine closely the extent of the stiffness of the neutron star EoS and the relative constraints on the upper bound on the speed of sound.

We now discuss the weighted tidal polarizability $\tilde{\lambda}$ as a function of the chirp mass $M{\cal M}$ varying the symmetric ratio $h$, as shown in Fig. 8.
Considering a binary neutron star system with masses $m_1$ and $m_2$ it easy to show that
\begin{equation}
m_1=\frac{{\cal M}}{2h^{3/5}}\left(1+\sqrt{1-4h}
\right), \qquad m_2=\frac{{\cal M}}{2h^{3/5}}\left(1-\sqrt{1-4h}
\right)
\label{m1,m2,l}.
\end{equation}
In this case the weighted $\tilde{\lambda}$ is rewritten as
\begin{equation}
\tilde{\lambda}=\frac{1}{26}\left[\left(1+12\frac{1-\sqrt{1-4h}}{1+\sqrt{1+4h}}  \right)\lambda_1({\cal M},h)+\left(1+12\frac{1+\sqrt{1-4h}}{1-\sqrt{1+4h}}  \right)\lambda_2({\cal M},h)
\right].
\label{aver-tital-2}
\end{equation}
We consider again the main three  cases (EoS/normal, EoS/minstiff, and EoS/max stiff) where for the intermediate region of the density we employ the MDI model with the slope parameter $L=95$ MeV. The three values of the symmetric ratio (0.25, 0.242, 0.222) correspond to the mass ratio $m_2/m_1$ (1.0, 0.7, 0.5) (for more details see also Ref.~\cite{Hinderer-10}). The uncertainty $\Delta\tilde{\lambda}$ in measuring $\tilde{\lambda}$ of the Advanced LIGO and the corresponding of the Einstein Telescope are also presented (see also Fig.~7 for more details). From Fig.~8 it is concluded that the upper bound of the speed of sound and consequently the  maximum mass configuration affects appreciable the chirp mass-weighted tidal polarizability dependence. This effect is more pronounced for chirp masses  ${\cal M} > 0.5 \ M_{\odot}$. In particular, for high values of ${\cal M}$, the Einstein telescope has the sensibility to distinguish the mentioned dependence. It is worth  pointing out also  the moderate, but visible dependence on the symmetric ratio $h$ in each case. This dependence is more effective for the EoS/max stiff case and for $h=0.222$, that is,  when the asymmetry of the two masses is very large.

In the present work we mainly deal with  the production of the maximum mass configurations and the possibility of their  observations. However, it is of interest to study also the maximum mass configuration effects on the corresponding radius (see also Fig.~3). This is shown in Fig. 9.  As one can see, the strong dependence of $\lambda$   on $R$ is common to all employed models.  This complex dependencies result from Eq.~(\ref{Love-1}) and, in particular, from the different  model-dependent neutron star configurations. Therefore, an accurate observation of the neutron star radius $R$ or tidal polarizability $\lambda$ will also help contribute to a determination of the EoS stiffness.

The schematic approach for neutron star matter proposed by Olson~\cite{Olson-2000} may be simple, however, the constraints based on the kinetic theory provide a useful {\it guide}  on the upper bound of the speed of sound. Actually,  there are many conjectures concerning the upper bound of the speed of sound in hadronic matter but, to our  knowledge, the bounds found from the relativistic kinetic theory are the only ones originated from an self-consistent treatment of hadronic matter. We thus consider that it is of interest  to present  the results using the  relativistic kinetic theory constraints on the EoS and to compare them with the  upper bounds suggested by other approaches as well as with a future relative observations.
In Fig.~10(a) we display the mass-radius dependence when the constraints on the upper limit of the speed of sound are taken into account according to the relativistic kinetic theory. We employ five values for the critical density $n_0$. The use of the upper bounds on the speed of sound, imposed by the kinetic theory, lead to values of $M_{{\rm max}}$ which do explain the recent observations, even for high values of the critical density $n_0$. The increase of $M_{{\rm max}}$ as well of the compactness parameter $\beta$ leads, as in the previous cases, to a significant increase on the values of the tidal polarizability $\lambda$ [see Fig.~10(b)]. This effect is more pronounced at high densities and, in particular, the $\lambda$ values are inside the sensibility of the Einstein Telescope.

\section{Concluding Remarks}
In the present work, we study in detail, the upper bound effects on the speed of sound on the EoS and the bulk properties of a neutron star. We focus on three extreme limits: the $v_s=c$, the  $v_s=c/\sqrt{3}$, and the limit imposed by the relativistic kinetic theory. Our investigation shows that: a) the constraint $v_s\leq c/\sqrt{3}$ on the EoS significantly reduces  the maximum mass of neutron stars ($M_{max} \leq 2. M_{\odot}$) for a class of nuclear models (non-relativistic as well as relativistic) and makes them inconsistent with the astrophysical observations of massive neutron stars. The constraints  $v_s \leq c$ and $\frac{v_s}{c}\leq \left(\frac{{\cal E}-P/3}{P+{\cal E}}\right)^{1/2}$ allow neutron star configurations with high neutron star masses, compatible with the recent observations, b) the tidal polarizability is also sensitive to the EoS and the relevant constraints introduces by the speed of sound.

We believe that the simultaneous measure of $M$ and $\lambda$ will help to better understand  the stiffness limit  of the equation of state. In particular, observations with  third-generation  detectors,  will definitely provide  constraints for the stiffness of the EoS at  high density. This is expected to provide more information related to the upper bound of the speed  of sound in hadronic matter. The accurate estimate of the upper bound of the speed  of sound in hadronic matter is greatly important for a consistent prediction of the maximum mass of a neutron star. The future detection and analysis of gravitational waves in binary neutron star systems is expected to shed light on this problem.


\begin{figure}
\centering
\includegraphics[height=8.5cm,width=10.5cm]{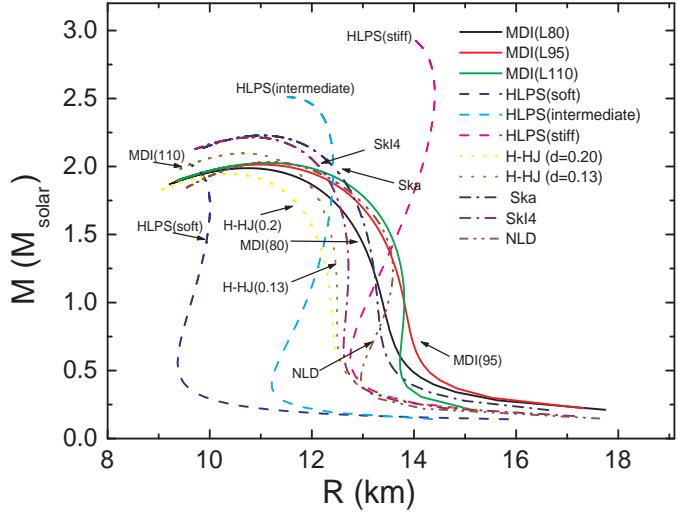}\\
\caption{ Mass-radius diagram for the equations of state used in the present work.   } \label{}
\end{figure}
\begin{figure}
\centering
\includegraphics[height=8.5cm,width=10.5cm]{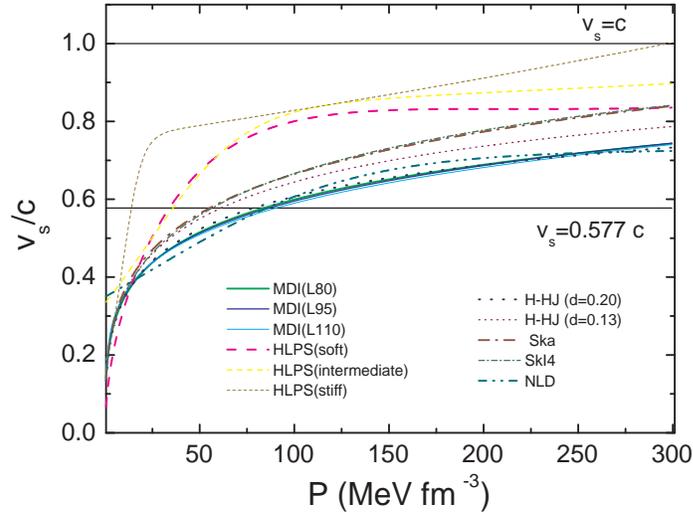}\
\caption{The speed of sound dependence on the pressure for the EoSs used in the paper. The two specific upper bounds considered in the present work
$v_s=c$ and $v_s=c/\sqrt{3}\simeq 0.577 c$ are also indicated.    } \label{}
\end{figure}
\begin{figure}
\centering
\includegraphics[height=8.5cm,width=10.5cm]{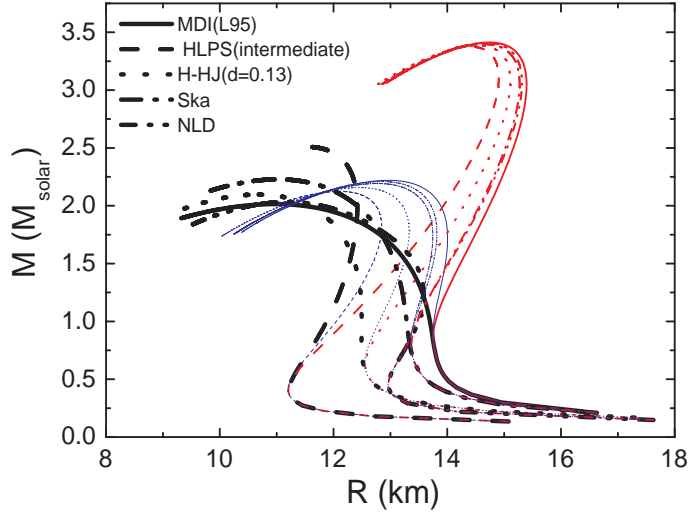}\
\caption{The mass-radius diagram for five  EoSs (EoS/normal case, line with thick width)  in comparison with the corresponding maximum mass configuration results of the EoS/minstiff case (line with medium width) and  and EoS/maxstiff case (line with thin width).} \label{}
\end{figure}

\begin{figure}
\centering
\includegraphics[height=8.5cm,width=10.5cm]{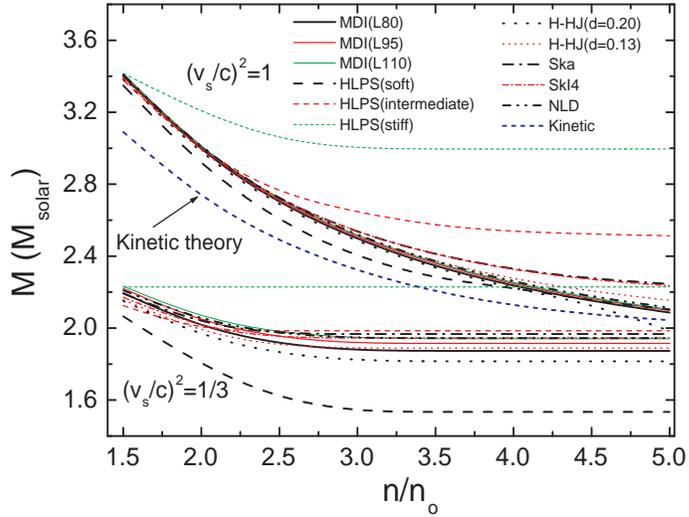}\
\caption{The maximum mass of neutron stars as a function  of the critical density $n_0$ for the two upper bounds for the speed of sound $v_s=c$ and $v_s=c/\sqrt{3}$. The case which corresponds to the upper bound for $v_s$, which is  taken from the kinetic theory, is also indicated.     } \label{}
\end{figure}

\begin{figure}
\centering
\includegraphics[height=8.5cm,width=10.5cm]{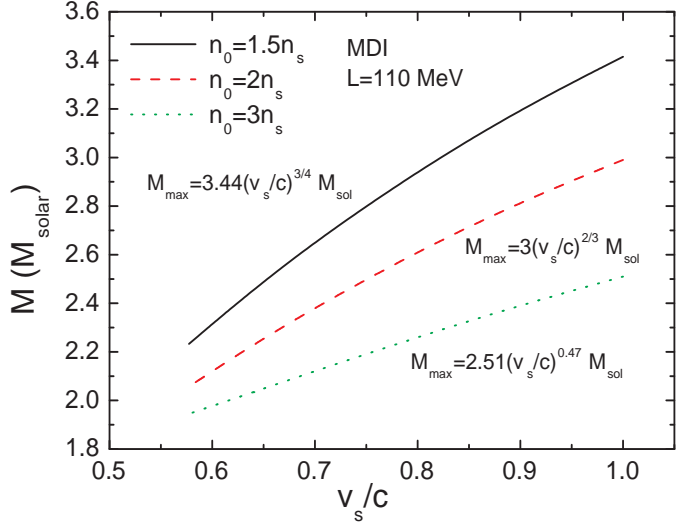}\
\caption{The maximum mass  as function of the upper bound of the speed of sound and for three different values of the critical density. The dependence is of the form $M_{\rm max}=a(v_s/c)^b M_{\odot}$. The values of the parameters $a$ and $b$, for each case, have
been selected by a least-squares fit method.  } \label{}
\end{figure}

\begin{figure}
\centering
\includegraphics[height=8.5cm,width=10.5cm]{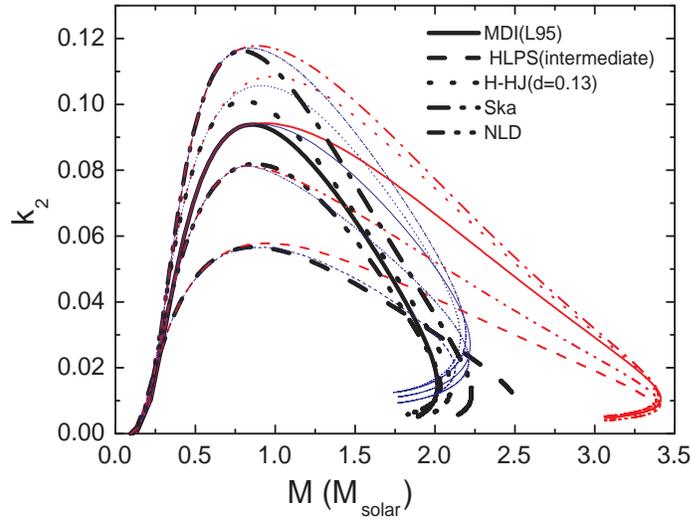}\
\caption{The tidal Love number $k_2$ as a function of the mass for the five selected EoS's (EoS/normal case)  in comparison with the corresponding maximum mass configurations results (EoS/minstiff and EoS/maxstiff cases). The notation is as in Fig.~3.   } \label{}
\end{figure}

\begin{figure}
\centering
\includegraphics[height=8.5cm,width=10.5cm]{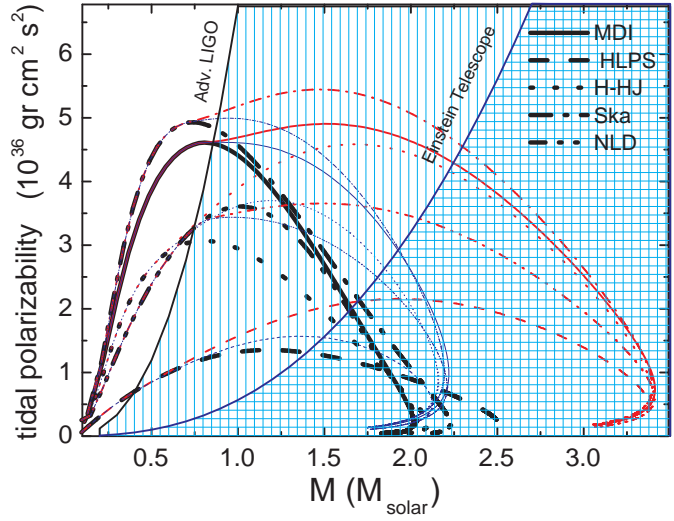}\\
\caption{The tidal polarizability $\lambda$  of a single neutron star as a function of the mass for the five selected EoS's (EoS/normal case)  in comparison with the corresponding maximum mass configurations results (EoS/minstiff and EoS/maxstiff cases). The notation is as in Fig.~3.  The ability detection region of the Advanced LIGO is the unshaded region and the corresponding of the Einstein Telescope  by the unshaded and light shaded region (see text for more details and also Ref.~\cite{Hinderer-10}).} \label{}
\end{figure}

\begin{figure}
\centering
\includegraphics[height=8.5cm,width=10.5cm]{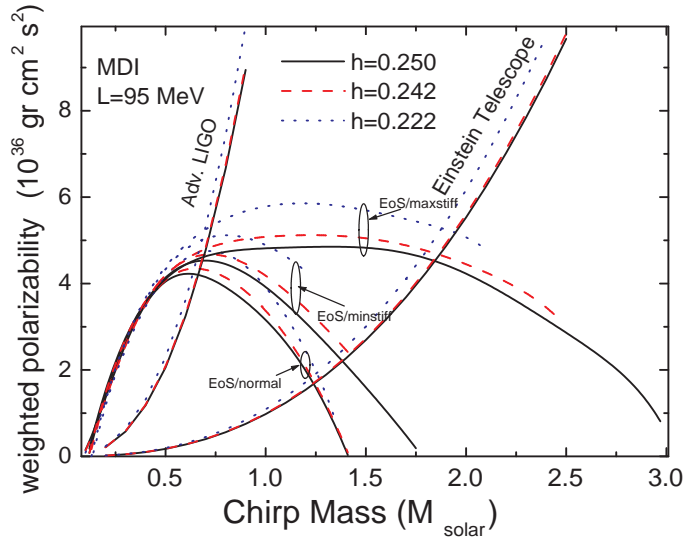}
\caption{The weighted tidal polarizability $\tilde{\lambda}$ as a function of the chirp mass ${\cal M}$ for various values of the symmetric ratio $\eta$ for the three considered cases (EoS/normal, EoS/minstiff, and EoS/max stiff) which correspond to the MDI (L95) EoS. The three values of the symmetric ratio (0.25, 0.242, 0.222) corresponds to the mass ratio $m_2/m_1$ (1, 0.7, 0.5) (see also Ref.~\cite{Hinderer-10} for comparison). The uncertainty $\Delta\tilde{\lambda}$ in the $\tilde{\lambda}$ measure of the Advanced LIGO and the corresponding of the Einstein Telescope  are also indicated (for more details see text  and also Ref.~\cite{Hinderer-10}).    } \label{}
\end{figure}
\begin{figure}
\centering
\includegraphics[height=8.5cm,width=10.5cm]{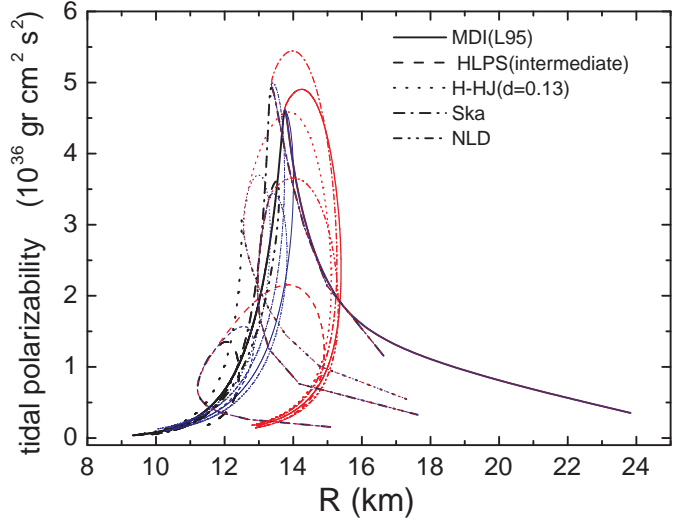}
\caption{The tidal polarizability $\lambda$ as  a function of the radius for the five selected EoS's in comparison with the corresponding maximum mass configurations results.    } \label{}
\end{figure}
\begin{figure}
\centering
\includegraphics[height=8.5cm,width=8.6cm]{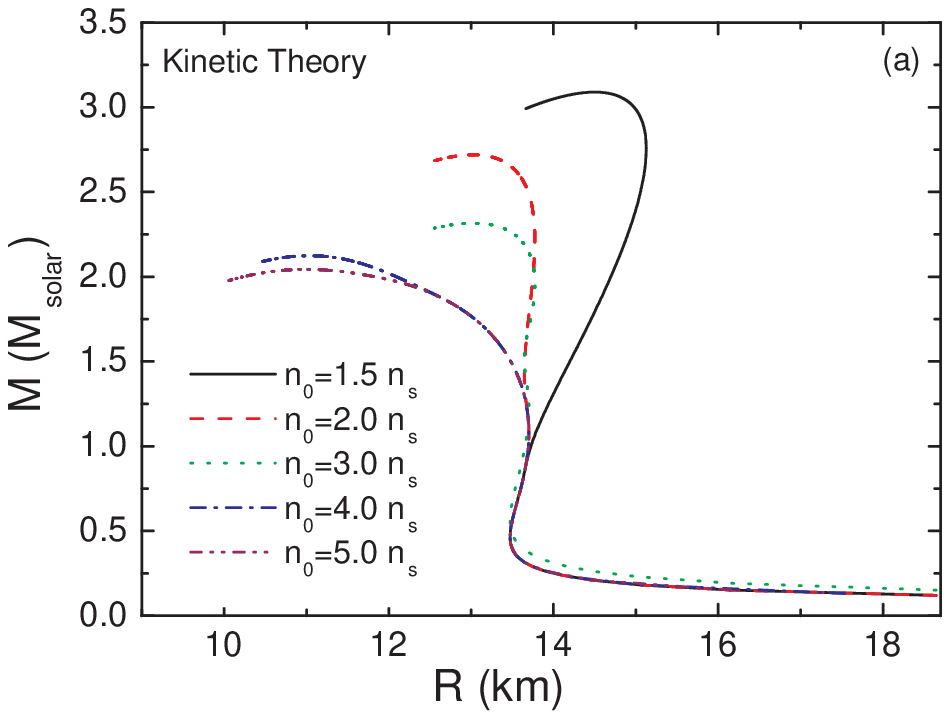}\
\includegraphics[height=8.5cm,width=8.6cm]{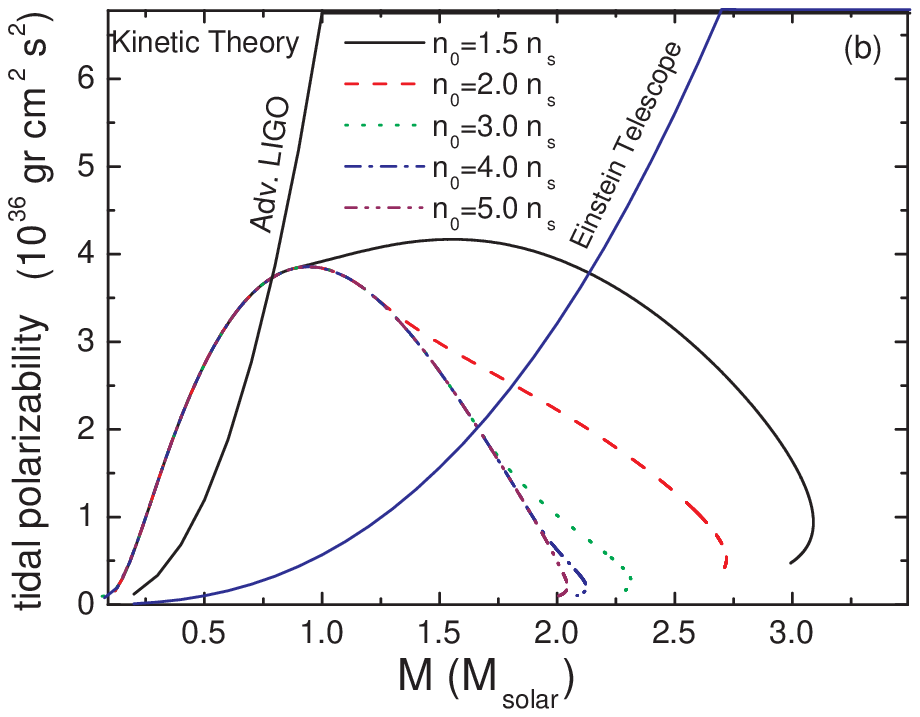}
\caption{(a) The mass-radius diagram which corresponds to the constraints for the relativistic kinetic theory in the $v_s$ and for various values of the critical density $n_0$ (see text for more details), (b) The $M$-$\lambda$ dependence from the kinetic theory in comparison with the observation abilities  (see text for more details). The uncertainty $\Delta \lambda$ in the $\lambda$ measure of the Advanced LIGO and the corresponding of the Einstein Telescope  are also indicated (see also Fig.~7).     } \label{}
\end{figure}

\end{document}